\documentclass[%
 aip,
 amsmath,amssymb,
 reprint,%
]{revtex4-1}

\usepackage{graphicx}
\usepackage{dcolumn}
\usepackage{bm}

\usepackage[utf8]{inputenc}
\usepackage[T1]{fontenc}
\usepackage{mathptmx}
\usepackage{etoolbox}
\usepackage{color}
\usepackage{hyperref}
\usepackage{booktabs}       
\usepackage{multirow}

\makeatletter
\def\@email#1#2{%
 \endgroup
 \patchcmd{\titleblock@produce}
  {\frontmatter@RRAPformat}
  {\frontmatter@RRAPformat{\produce@RRAP{*#1\href{mailto:#2}{#2}}}\frontmatter@RRAPformat}
  {}{}
}%
\makeatother
\begin{document}

\preprint{AIP/123-QED}

\title[PANNA 2.0]{PANNA 2.0: Efficient neural network interatomic potentials and new architectures}

\author{Franco Pellegrini}
\email{pellefra@sissa.it}
\affiliation{Scuola Internazionale Superiore di Studi Avanzati, Trieste, Italy.}

\author{Ruggero Lot}
\affiliation{Scuola Internazionale Superiore di Studi Avanzati, Trieste, Italy.}

\author{Yusuf Shaidu}
\affiliation{Department of Physics, University of
California Berkeley, Berkeley, California 94720, United
States\\}
\affiliation{Materials Sciences Division, Lawrence
Berkeley National Laboratory, Berkeley, California 94720,
United States\\}
\affiliation{Scuola Internazionale Superiore di Studi Avanzati, Trieste, Italy.}

\author{Emine K\"{u}\c{c}\"{u}kbenli}
\affiliation{Nvidia Corporation, Santa Clara, CA, USA}
\affiliation{John A. Paulson School of Engineering and Applied Sciences, Harvard University, Cambridge, Massachusetts 02138, USA}
\email{ekucukbenli@nvidia.com}

\date{\today}

\begin{abstract}
We present the latest release of PANNA 2.0 (Properties from Artificial Neural Network Architectures), a code for the generation of neural network interatomic potentials based on local atomic descriptors and multilayer perceptrons. Built on a new back end, this new release of PANNA features improved tools for customizing and monitoring network training, better GPU support including a fast descriptor calculator, new plugins for external codes and a new architecture for the inclusion of long-range electrostatic interactions through a variational charge equilibration scheme.
We present an overview of the main features of the new code, and several benchmarks comparing the accuracy of PANNA models to the state of the art, on commonly used benchmarks as well as richer datasets.
\end{abstract}

\maketitle

\section{Introduction}

In recent years, machine learning (ML) based approaches have been successfully applied to numerous problems, spanning from image and natural language processing to many areas of physics.
Within the field of atomistic simulations, several approaches have been presented to leverage ML for the accurate prediction of molecular and material properties. In particular, one of the main goals has been the fast computation of energies and forces, leading to the creation of ML-based interatomic potentials (MLIPs), able to achieve the accuracy of \textit{ab initio} methods on selected systems, for a fraction of the cost.
While we refer the reader to the many available reviews~\cite{Behler16,Unke21,Kulik22} for an exhaustive presentation of ML approaches in material science, we will briefly present here the main flavors of MLIPs present in literature, to provide a context for implementations within PANNA.

Most MLIPs are based on two approximations: i) the possibility to write the total energy of a system as a sum of atomic contributions, ii) spatial locality. This allows to roughly break down the problem into two parts: defining---or learning---a description of a local atomic environment, and learning a function to map the descriptor to the local energy.
The requirement for invariance with respect to translations, rotations, and permutations of the atoms is enforced exactly by either  invariant descriptors or equivariant network architectures.

The earlier methods in the field to describe the local environment relied on fixed descriptors, e.g. Behler-Parrinello~\cite{BP,ANI} (BP) type descriptors sample the two- and three-body distribution function with local sampling functions; while the Smooth Overlap of Atomic Positions~\cite{De16} (SOAP) relies on spherical harmonics to obtain a rotationally invariant description of a power of the smoothed atomic density.
In these cases, ML was limited to the mapping of descriptors to atomic quantities, which relied, for example, on multilayer perceptrons~\cite{BP,ANI} (MLPs), or kernel methods, as in the case of the Gaussian approximation potential~\cite{GAP} (GAP).
These and similar approaches have been shown to achieve chemical accuracy in a host of different systems~\cite{Deringer17,Shaidu21,Artrith12, Jain21}, typically given a ground truth of a few thousands configurations to train on.
A limit to the generalization capacity for a given number of training points, however, is related to the architectural bias of the approach, depending both on the ML model and the descriptors. Indeed, more advanced descriptors like the Atomic Cluster Expansion~\cite{Drautz} (ACE) were shown to obtain lower generalization errors with the same amount of data, even when the fitting was done through a simple linear model~\cite{ACElinear}.

In search for a better architectural bias, more advanced message passing~\cite{Gilmer} (MP), interaction layers~\cite{Lubbers18}, continuous filter convolution~\cite{SchNet}, or graph neural networks (GNN) were introduced: some using vectors, angles and other geometric information to define the node functions~\cite{DimeNet, GemNet, NewtonNet} and some promoting the states of the networks themselves to equivariant entities based on vectors~\cite{EGNN} or a basis of spherical irreducible representations~\cite{NequIP, MACE, Allegro}.
While the distinction between (learned) descriptor and ML model becomes blurred in these cases~\cite{BOTNet}, it has been clearly shown that the bias imposed by these architectures can lead to better generalization accuracy with the same amount of data. However, several layers of message passing can lead to very large effective receptive fields for each atom, and even when this can be avoided, each layer typically involves the use of several MLPs, leading to a larger overall computational cost. While the scaling with respect to  \textit{ab initio} is still favorable, this increased cost renders the earlier MLP approaches still valuable, especially for applications where sufficient data can be generated.

In this work, we present a new implementation of PANNA~\cite{PANNA1} (Properties from Artificial Neural Network Architectures), version 2.0, a package for the creation and deployment of MLIPs based on local descriptors and Multilayer perceptrons. Several packages have been proposed to train this type of networks, e.g. DeepMD~\cite{DeepMD}, AENet~\cite{AENet}, AMP~\cite{AMP}, TorchANI~\cite{TorchANI}, and SIMPLE-NN~\cite{SIMPLENN}. 
These packages are written in different languages (FORTRAN, Python), over different back ends (TensorFlow, PyTorch), and while some of them are based on input files, others provide an API to be called from user-written code. They are all based on atomic MLPs, but they support different descriptors, from the original BP~\cite{BP} to ones with modifications~\cite{ANI, Artrith17}; supporting different training features, network customization, learning schedules, ensemble approaches and so on. With this latest version of PANNA we hope to enrich this landscape, where variety allows more options to be explored and more needs to be met.

With respect to the previous version, the PANNA suite has been entirely rewritten to be compatible with the newest versions (2.x) of the TensorFlow~\cite{TF} framework. While supporting all the features of the earlier version, the code has been optimized to run on GPU, and it supports new features, such as the computation of descriptors during training, and a new architecture to handle long range electrostatic interactions. PANNA  is written in Python, and it can be simply run by supplying appropriate input configuration files. It includes several tools to customize and monitor the training, both through a graphical interface and from command line, as well as tools to import and export data from and to different external codes. Finally, PANNA models can be exported to run molecular dynamics (MD) directly in popular packages such as LAMMPS~\cite{LAMMPS} and ASE~\cite{ASE}, or to even more codes through an interface with OpenKIM~\cite{KIM}.
The PANNA code is released under an MIT license, and it can be downloaded at Ref.~\citenum{PANNAurl}. A thorough documentation, including a list of all input file keywords and several tutorials on how to run different example cases, is available at Ref.~\citenum{PANNAdoc}. 
In the following, we will present the main features of the code and the underlying theory in Sec.~\ref{sec:code}, and we will report benchmarks on accuracy on different systems, speed and data scaling in Sec.~\ref{sec:benchmark}.

\section{The Implementation}\label{sec:code}
The core of PANNA 2.0 is based on the creation of fixed-size atomic descriptors as inputs to MLPs for the computation of atomic energies, summing to the total energy of a system. Distinct architectures can be defined for each atomic species, and weights are shared between all atoms of the same species.
The training procedure consists in optimizing the MLP parameters to match the energy, and forces as its derivatives, on known configurations. This optimization is performed by minimizing a \textit{loss function} of the error, through stochastic gradient descent on small sets of examples known as \textit{mini-batches}.
In the next sections, we will highlight the options available in PANNA 2.0 for each step of this training procedure, and we will discuss specifically a new architecture that models long-range electrostatic interactions.

\subsection{General structure}\label{ssec:structure}
A typical MLIP training pipeline starts with the reference energies and forces being computed with density functional theory (DFT) or some other reference approach. In PANNA, we offer tools to convert the output of codes such as Quantum ESPRESSO~\cite{QE}, VASP~\cite{VASP}, USPEX~\cite{USPEX} and LAMMPS~\cite{LAMMPS} to a simple human readable format. This format is completely documented, so that a user can easily create a new converter from a different code.

In the next step of the pipeline, features or descriptors need to be computed for each atom. PANNA offers two ways of computing the descriptors: they can either be precomputed for the whole dataset, or they can be computed during training on-the-fly. The first option is computationally advantageous as examples are typically reused multiple times throughout the training, and since the descriptor is fixed it is possible to create it once and for all. This can however pose the problem of storing a large amount of data (especially when the derivatives of the descriptor are needed to compute the forces), and reading the data multiple times from storage (if they do not fit in the working memory). For this reason, we also offer the second option which, while more computationally expensive, makes it feasible to train on very large datasets (see Sec.~\ref{ssec:Carbon}). This option is also convenient for performing quick training cycles with various descriptor types and shapes for testing purposes without having to read/write large files.
PANNA natively includes routines to compute the standard BP descriptor~\cite{BP} and a modified version (mBP) as detailed in the previous version of PANNA~\cite{PANNA1}. For precomputed descriptors, the binary format used for storage is carefully documented such that descriptors computed with external routines can be adapted to the PANNA pipeline.

PANNA currently implements MLP type networks. The general equation for the architecture is as follows:
\begin{equation}
    a^{l}_i = \sigma\left(\sum_{j=1}^{n_{l-1}}w^l_{ij}a^{l-1}_j+b^l_i\right),
\end{equation}
where $a_i$ is a node of layer $l$, $w$ and $b$ are weights and biases---the parameters of the network---$\sigma$ is a nonlinear function, and $n_l$ is the number of nodes in layer $l$ (input is considered as layer 0).
In PANNA, users can easily define the desired architecture, on a per-species basis, by specifying the size of the layers. The last layer is typically of size one for a single output for energy, but see Sec.~\ref{ssec:longrangeTheory} for a different case. The activation function $\sigma$ can be chosen for each layer as Gaussian, rectified linear unit (ReLU), or hyperbolic tangent, besides the linear function which is typically used for the output. Since PANNA is built on TensorFlow, 
the supported activation functions can be easily extended if desired, to the vast list supported by this framework.

The remaining elements of a training pipeline are the loss and learning schedule. The loss function in PANNA is made up of contributions coming from the energy and the forces errors, with an adjustable relative factor. For both, users can choose between a quadratic and exponential function of the difference between the computed and expected values, and whether to consider per configuration or per atom quantities.
A further regularization term can be added to the loss function as the sum of the absolute value (L1) or square (L2) of all weights, with a chosen prefactor.

The optimization of weights and biases, i.e. training, is finally performed on mini-batches of chosen size, modifying  $w$ and $b$ according to the gradient of the loss through the Adam~\cite{Adam} algorithm, with a learning rate that can be chosen as constant or exponentially decaying. As in the case of activation functions, TensorFlow backend provides PANNA with multiple options for optimizers. Additionally, freezing weights of selected layers for selected species is allowed to facilitate fine tuning or transfer learning studies.

During training, several tools within PANNA can be used to monitor the progress of the optimization. From the command line, users can decide to monitor the loss components or a figure of merit such as the root mean square error (RMSE) or mean absolute error (MAE). TensorFlow provides a graphical interface, TensorBoard, a browser based visualization tool. TensorBoard allows PANNA users to visualize loss components as well as other figures of merit, along with the evolution of the distribution of the weights and biases throughout the training. 
Moreover, once per epoch (or at a chosen frequency) the model can also be automatically evaluated on a validation set on-the-fly, to keep track of the generalization capacity or for decision of early-stopping. 

After the model is trained, it can be stored as a checkpoint, and PANNA's inference tool can be used to assess its performance on a testset. The model can also be exported to a format usable in external MD codes, e.g.  in LAMMPS~\cite{LAMMPS} thanks to the plugin included in the PANNA (now improved with OpenMP parallelization), or with many other MD packages supported via OpenKIM~\cite{KIM}. Alternatively, the internal checkpoint format, can be imported in ASE~\cite{ASE} through the calculator included in PANNA. The performance of this new plugin is tested in Sec.~\ref{ssec:Carbon}. Extension of PANNA potentials for modern MD  packages such as the differentiable JAX-MD~\cite{jaxmd} will be supported in the next version.

\subsection{Long range interactions}\label{ssec:longrangeTheory}
PANNA 2.0 supports a new method to address long-range electrostatic interactions. Most MLIP schemes rely an a locality approximation. While this is often safe to do in neutral systems, and might work for shielded charges, it is bound to fail when electrostatics plays a role in a range larger than the effective cutoff radius or in charged systems.

In recent years, to address this challenge, methods that couple a local network predicting atomic electronegativity with a system-wide charge equilibration scheme have been proposed ~\cite{CENT,Behler4G, QRNN, Shaidu23}. Ref.~\citenum{CENT} only deals in the electrostatic part, and Ref.~\citenum{Behler4G} proposes to employ two different networks (one dependent on the other). The implementation within PANNA is based on Ref.~\citenum{Shaidu23} and relies on a single network to predict the coefficients for a Taylor expansion of the energy in local charges, up to the second order. This allows PANNA to compute the total energy, including electrostatics, by evaluating a single network and solving a linear charge equilibration system. 

We leave a more in-depth theoretical exposition of this approach to be presented elsewhere ~\cite{Shaidu23} and focus on the implementation changes it brings. In this approach there are 3 different outputs for each atom: the coefficients of the Taylor expansion corresponding to local energy, electronegativity and atomic hardness (or their corrections with respect to a reference). These outputs are fed to a newly introduced network layer where the charge equilibration optimization takes place. The final outcome is the predicted local charges and the total energy. While the ground-truth local charge information can be used during training, it is not strictly necessary. This allows the use of many publicly available datasets where no local charge decomposition information has been stored. Interestingly, we find that the absence of reference local atomic charges (an approximate partition in many cases) can even improve the ability of the network to predict total energies and forces (see Sec.~\ref{ssec:longrangeExp}).

\section{Benchmarks}\label{sec:benchmark}

\subsection{rMD17}\label{ssec:rMD17}

\begin{table}[t]
\caption{
Mean absolute error in energy (meV on the whole molecule) and forces (meV/\AA\ per component) of different models trained on 1000 configurations from each molecule in the rMD17 dataset~\cite{rMD17}.
The ANI results are taken from Ref.~\citenum{ACElinear}, where ANI was either trained from scratch, column ``ANI (rand)'' or starting from a pretrained model, column ``ANI (pre)''.
In the last column we report the state of the art (SOTA), i.e. the best result found for any model, giving priority to the force error, and the respective reference.
}
\centering
\label{tab:rmd17}
\resizebox{.48\textwidth}{!}{
\begin{tabular}{@{}llccc|l@{}}
\toprule
&& PANNA& ANI~\cite{ANI2} (pre) & ANI~\cite{ANI2} (rand)&{SOTA}\\
\midrule
\multirow{2}{*}{\textbf{Aspirin}}
& E & 10.6& 16.6&25.4&2.2~\cite{MACE}\\
& F & 32.9& 40.6&75.0&6.6\\
\midrule
\multirow{2}{*}{\textbf{Azobenzene}}
& E & 5.8&15.9&19.0&1.2~\cite{Allegro}\\
& F & 18.4&35.4&52.1&2.6\\
\midrule
\multirow{2}{*}{\textbf{Benzene}}
& E & 1.0& 3.3&3.4&0.3~\cite{Allegro}\\
& F & 5.4& 10.0&17.4&0.2\\
\midrule
\multirow{2}{*}{\textbf{Ethanol}}
& E &2.9 &2.5&7.7&0.4~\cite{MACE,Allegro}\\
& F & 16.5& 13.4&45.6&2.1\\
\midrule
\multirow{2}{*}{\textbf{Malonaldehyde}}
& E & 4.0&4.6&9.4&0.6~\cite{Allegro}\\
& F & 24.3& 24.5&52.4&3.6 \\
\midrule
\multirow{2}{*}{\textbf{Naphthalene}}
& E & 3.0& 11.3&16.0&0.2~\cite{Allegro}\\
& F & 13.2& 29.2&52.2&0.9\\
\midrule
\multirow{2}{*}{\textbf{Paracetamol}}
& E & 6.3& 11.5&18.2&1.3~\cite{MACE} \\
& F & 22.0& 30.4&63.3&4.8\\
\midrule
\multirow{2}{*}{\textbf{Salicylic acid}}
& E & 4.1& 9.2&13.5&0.9~\cite{Allegro} \\
& F & 19.4& 29.7&53.0&2.9\\
\midrule
\multirow{2}{*}{\textbf{Toluene}}
& E &3.9& 7.7&12.6&0.5~\cite{MACE}\\
& F & 15.9& 24.3&52.9&1.5\\
\midrule
\multirow{2}{*}{\textbf{Uracil}}
& E & 2.4& 5.1&8.4&0.6~\cite{Allegro}\\
& F & 13.7& 21.4&44.1&1.8 \\
\bottomrule
\end{tabular}
}
\end{table}

The rMD17 benchmark set consists of configurations of 10 small organic molecules, with energies and forces computed in DFT with a tight convergence threshold~\cite{rMD17}. In recent years it has been commonly used to benchmark the data efficiency of MLIPs, specifically restricting the training set to a budget of 1000 randomly selected configurations per molecule. While this has shown the high data efficiency of equivariant GNN, the only typically reported MLP with BP type descriptors (ANI~\cite{ANI}) seemed to fall considerably behind. We demonstrate here the performance of PANNA with an equivalent BP type network for comparison.

The computational details are as follows: An mBP type descriptor~\cite{PANNA1} with a maximum cutoff of 5\AA was used with 24 radial bins. For the angular part, 8 bins for the average radius, and 8 for the angles, i.e. a total of 64 angular bins were used. Considering 2, 3 or 4 species, this resulted in descriptors of size 240, 456, and 736, respectively. We then trained networks with 2 hidden layers of size 256 and 128, training for $10^6$ steps with learning rate $10^{-4}$, and then reducing it to $10^{-6}$ over a further $10^6$ steps. We employed quadratic loss with a force cost of 1 and a small $10^{-5}$ L1 regularization.

The validation MAE in energy and force components is reported in Table~\ref{tab:rmd17}, alongside the results of ANI and the state-of-the-art selected from literature considering all architectures including kernel methods. While the supremacy of equivariant GNN remains unreachable for this kind of networks (all SOTA values are found to be from various equivariant GNNs), errors from PANNA are considerably lower than ANI, for almost all molecules. This is particularly interesting as the most performant ANI network reported thus far in these benchmarks were obtained by retraining a pretrained network (``ANI (pre)'' in Table~\ref{tab:rmd17}) for increased accuracy ~\cite{ACElinear}. Our results show that even for the same method and data, the differences in training and implementation can have a major impact on the final model quality, strenghtening the argument that well-written and well-documented MLIP generation packages are needed to reach a consistent quality of applications in the literature.

\subsection{Dataset scaling: Carbon}\label{ssec:Carbon}
To further assess the generalization capacity of our model, as a function of the size of the dataset, we consider a more challenging problem: a dataset with more than 60000 configurations of various allotropes of Carbon, created in a recent study using an evolutionary algorithm and the previous version of PANNA~\cite{Shaidu21}. The datasets consists mostly of configurations with 16 or 24 Carbon atoms, and a few larger (200 atoms) configurations; it includes configurations under high pressure, snapshots of high temperature MD and highly defected configurations (see Ref.~\citenum{Shaidu21} for a complete description and construction procedure).

\begin{figure}
    \centering
    \includegraphics[width=\linewidth]{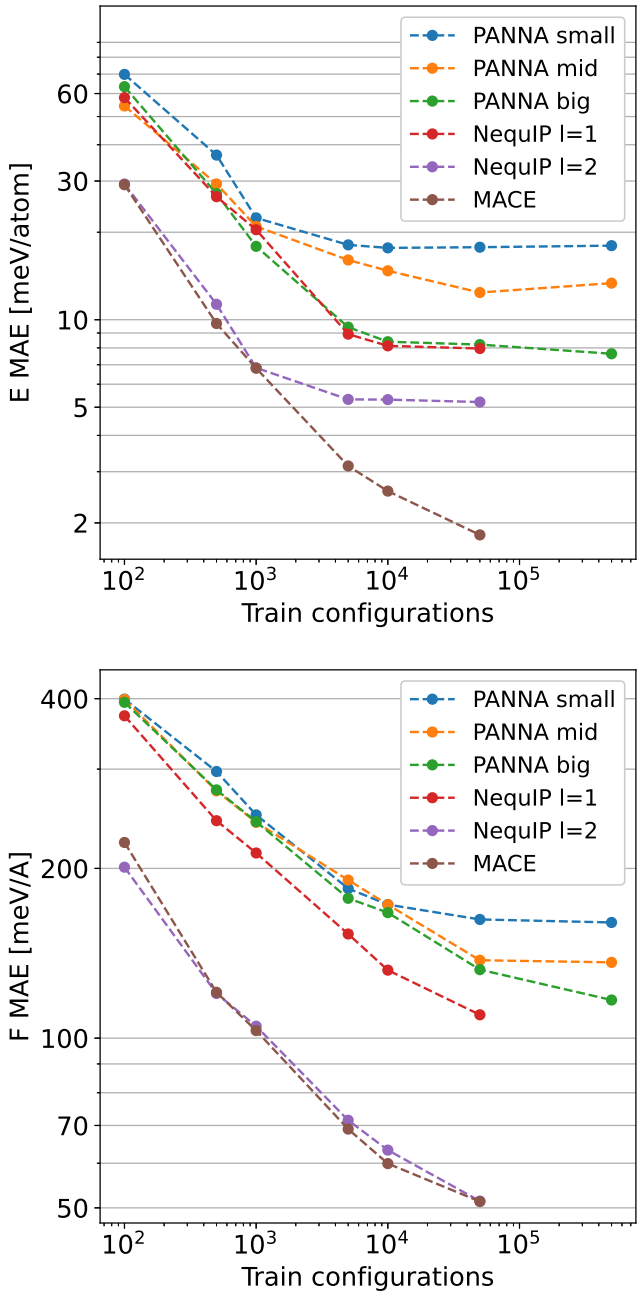}
    \caption{Scaling of the mean absolute error in energy per atom (top) and forces per component (bottom) for different models, as a function of the size of the training dataset. See the main text for further details on the training, including the definition of the three architectures PANNA small, mid and big.}
    \label{fig:Carbon}
\end{figure}

We split the dataset in 50000 randomly chosen configurations for training and we save the rest for validation. In order to generate well sampled training datasets of different sizes, we employ a farthest point clustering algorithm: we consider the cosine fingerprint distance as defined in Ref.~\citenum{Shaidu21}, we then start from a set of a single configuration (the lowest in energy) and progressively add the configuration that is farthest. In this way we generate datasets ranging from 100 to the whole 50000 configurations. We sample 1000 configurations from the validation set with the same approach.

For PANNA, we employ a mBP descriptor with 5~\AA\ cutoff, 24 radial bins for the two body and 8 radial and 16 angular bins for the three body terms, for a total size of 152. We train 3 different networks, a small one with two layers of sizes 64 and 32, a middle one with two layers of 256 and 128 nodes and a large one with three layers of sizes 1024, 512 and 256. Networks are trained on batches of 10 examples with a starting learning rate of $10^{-4}$ for a number of steps ranging from 100000 for the smallest dataset, to 6 millions for the largest one, after which a further quench to a learning rate $10^{-6}$ for 1 million steps is performed for all datasets larger than 1000 points. 

As reference state-of-the-art models we consider NequIP~\cite{NequIP} and MACE~\cite{MACE}. We train both on all datasets relying on default parameters as needed: for NequIP we consider two models,  with $\ell=1$ and $\ell=2$, both with 4 interaction blocks, 32 features and a radial network with 8 basis functions, and 2 hidden layers of size 64. For MACE we use the standard setup with 128 even scalars and 128 odd $\ell=1$ irreps. We use the same cutoff of 5~\AA\ for all models although the effective receptive field will be larger depending on the number of layers of GNNs.

For all networks we train until loss convergence. An important remark needs to be made about the training dynamics: There is an apparent trade-off between energy and force components in the loss especially close to convergence and for the case of very large datasets. To tackle this, MACE training schedule implements stochastic weight averaging (SWA)~\cite{SWA} and  increases the energy weight for the loss in the lass 20\% of the training. We find that in the case of NequIP and PANNA where a standard non-averaged optimizers such as Adam~\cite{Adam} with fixed energy and loss weighing is used, the energy loss decreases very minimally even when increasing the relative weight of it in the loss component, even in long training scenarios, suggesting the SWA and energy re-weighing during training can be valuable.
Overall we observe the force error to be more stable for all networks (given also the larger number of force data), and because the trade-off is hard to quantify for each model, we focus more on the force error during analysis.

Without significant hyperparameter tuning we do not expect these to be the best possible networks (including the one of PANNA), yet informative for a typical user experience. Fig.~\ref{fig:Carbon} shows the MAE of the error in energy per atom and in forces per component. Overall, equivariant models, especially with high $\ell$ orders, perform better. As mentioned, the failure to improve the energy error for the larger datasets is visible for PANNA and NequIP. Among the PANNA models, we can see all models obtain similar results for small datasets, and as the dataset becomes larger the smaller model seems to reach its capacity and its performance drops.

Considering that MLP architectures such as BP networks are much more computationally affordable due to simple underlying tensor algebra compared to irrep algebra of equivariant GNNs, it would be desirable to find a strategy to overcome for their data-inefficiency. Here we show a potential workaround with a data augmentation experiment. Starting with the largest training dataset, for each example we create 10 copies by perturbing the atomic positions randomly with a small Gaussian noise of standard deviation 0.075\AA. We then take the best MACE network and use it to compute energy and forces, obtaining a new dataset of half a million configurations, at a fraction of the DFT cost. Retraining PANNA models on this new larger dataset shows that a further improvement in accuracy can be obtained for large enough models. This ``knowledge distillation'' procedure is well known in the ML literature~\cite{Hinton15} and here too it proves to be a potential approach to keep less data-efficient models viable at a reduced cost.

\begin{table}[t]
\caption{
Time per step per atom to run a Langevin MD on a small Carbon cell with different codes, invoked through ASE on GPU.
}
\centering
\label{tab:Cspeed}
\begin{tabular}{lcccccc}
\toprule
& PANNA & PANNA & PANNA & NequIP & NequIP & MACE\\
& small & middle & big & $\ell=1$ & $\ell=2$ &\\
\midrule
Time [ms]& 0.78 & 0.79 & 0.79 & 3.32 & 4.88 & 3.09\\
\bottomrule
\end{tabular}
\end{table}

Lastly, we consider the computational performances of these potentials when used for inference: we take one of the configurations from the dataset with 16 atoms and perform 1000 steps of Langevin dynamics at a temperature of $300$~K with  ASE~\cite{ASE} on a A100 GPU, discarding the first few steps which typically require extra setup time, not representative of the speed of the codes. To judge the natural scalability of different algorithms we refrain from using any specialized optimization techniques such as CUDA based featurization library implemented in TorchANI \cite{TorchANI} for BP networks, as similar ones for tensor product in the equivariant GNNs are not yet widely available. Table~\ref{tab:Cspeed} reports the time per step per atom of each code, as an average of 5 repetitions. We note that due to small system size, these values should be taken as upperbound, since GPU utilization is far from optimal for these system sizes. Hardware specific strategies such as multi-instance or multi-process (MIG or MPS) GPU features can potentially bring significant improvements to GPU use. These and further optimization opportunities of BP networks will be employed in future versions of PANNA. Nevertheless, without specific strategies, a raw comparison of algorithms on similar grounds show that PANNA is consistently faster, as expected from a simpler architecture. It is noteworthy that larger PANNA models are not slower, as the computational bottleneck is in the calculation of the descriptor, hence, acceleration CUDA libraries as mentioned earlier, or descriptors with lesser computational load such as ACE \cite{Drautz} can bring further speedup. 

\subsection{Long range: NaCl clusters}\label{ssec:longrangeExp}
\begin{figure}[!hbt]
    \centering
    \includegraphics[width=\linewidth]{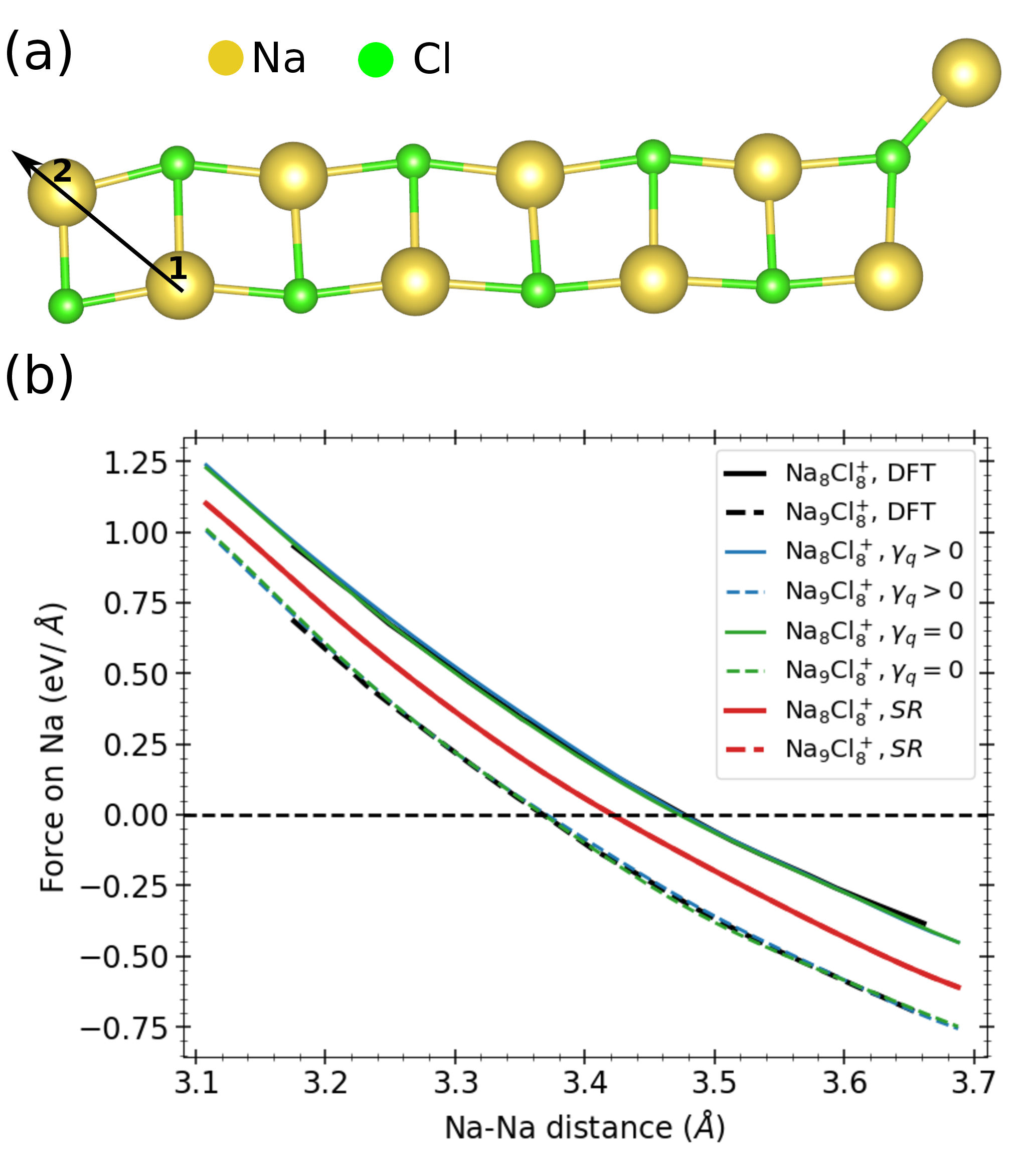}
    \caption{Comparison of energy and forces between MLIPs and DFT. (a) Atomic structure of Na$_9$Cl$_8^+$. 
    (b) Projected force on Na atom 1 in the direction of the arrow shown, as a function of the distance between Na atom 1 and 2. 
    }
    \label{fig:PES}
\end{figure}

In this section, we demonstrate the long range electrostatic approach implemented in PANNA on charged sodium chloride clusters. The training set is obtained from Ref.~\citenum{Behler4G} and comprises of configurations of Na$_9$Cl$_8^+$---shown in Fig.\ref{fig:PES}a---and Na$_8$Cl$_8^+$, obtained by removing the Na atom in the rightmost corner of Fig.~\ref{fig:PES}a. Each cluster has a total charge of +1. Here we compare the accuracy of long range model within PANNA with that reported in Ref.~\citenum{Behler4G}.

The MLIP is constructed with mBP atomic environment descriptors of size 45, two hidden layers each with 15 nodes and an output layer with 3 nodes. As explained in Sec.~\ref{ssec:longrangeTheory}, in PANNA reference charges can be either used as an extra target in the loss function or omitted: we present here one model with ($\gamma_q>0$) and one without ($\gamma_q=0$) this extra loss. We also compare with the PANNA model in the absence of long range electrostatics (SR for short range), with the same architecture but only predicting energy and forces.

In Table~\ref{NaCl_table}, RMSE in charges, energy and forces for PANNA models are compared with the results obtained in Ref.~\citenum{Behler4G} with and without long range interactions, denoted as 2G- and 4G-HDNNP, respectively. The PANNA model with long range electrostatics reaches the lowest RMSE in energy and forces irrespective of the use of atomic charges as target. It is reassuring for verification reasons that, as a baseline, The PANNA model without electrostatics attains similar RMSE in energy and forces to 2G-HDNNP~\cite{Behler4G} that also omits this contribution.

 \begin{table}
    \caption{Training and validation RMSE of different quantities for the sodium chloride cluster with a total charge of +1 for long range (LR) and short range (SR) models, and models from Ref~\citenum{Behler4G} (2G, 4G).}
     \begin{tabular}{lllccc}
     \toprule
     Model&&&Charge&Energy&Force\\
          &&&[me]&[meV/atom]&[meV/\AA]\\
     \midrule
     SR\\
      && train  & ~-~ & 1.6 & 49\\
      && test  & ~~ & 1.6 & 50 \\
     \midrule
     LR\\
      $\gamma_q>0$&& train  & 11.5 & 0.4 & 17\\
        && test  & 11.8 & 0.4 & 18\\
      $\gamma_q=0$&& train  & 237.3 & 0.3 & 17\\
        && test  & 238.9 & 0.3 & 18\\
     \midrule
     2G&&train & ~-~ & 1.7 & 58 \\
       &&test    & ~-~ & 1.7 & 57 \\   
     4G&&train & ~15.9 & 0.5 & 32 \\
       &&test    & ~15.8 & 0.5 & 33 \\
     \bottomrule
     \end{tabular}
    \label{NaCl_table}
    \end{table}

We examine in further detail the performance on the potential energy surface of these systems by computing the energy and force acting on a Na atom, indicated by 2 in Fig~\ref{fig:PES}a, when moved along the arrow depicted in the same figure. 
Fig.~\ref{fig:PES}(b) shows the force on the Na atom projected along the direction shown by the arrow. The DFT results are obtained from Ref.~\citenum{Behler4G}. As expected, for the PANNA model without electrostatics, we obtain similar trends to those reported in Ref.~\citenum{Behler4G} for the 2G-HDNNP, where the equilibrium distances for Na$_8$Cl$_8^+$ and Na$_9$Cl$_8^+$ are the same. Instead, model with long range electrostatics accurately reproduces the DFT forces for both Na$_9$Cl$_8^+$ and Na$_8$Cl$_8^+$, with and without regressing against reference charges. We note that the forces as a function of distance is smooth, suggesting stable dynamics and possibility of obtaining energy differences through integrating the forces if needed.

\section{Conclusion}\label{sec:conclusion}
We have given a brief overview of PANNA 2.0, the latest version. Besides the support for the new version of the Tensorflow back end---a needed upgrade to run on newer hardware where previous versions are becoming increasingly harder to obtain---this new version features several improvements aimed at simplifying the training procedure for the end user. Removing the need to precompute descriptors simplifies the exploration of new parameters, or training on very large datasets; new figures of merit and validation on-the-fly make it easier to monitor the optimization in real time. Importantly, PANNA 2.0 introduces support for long-range electrostatics, which opens the possibility to tackle charged systems that were not accessible before.

Moreover, we have shown in a series of benchmarks that while the PANNA models are not as data efficient as the newest equivariant GNN architectures, they can be more accurate than what previously reported for similar models, and they do show an accuracy-scaling power law dependence on the size of the dataset that is comparable to some equivariant models. We have also proposed the ``knowledge distillation'' scheme to employ the more data efficient networks to extend the training set for the less data efficient ones. Paired with fast MD plugins, these results point towards a possibility where simple architectures like PANNA can become the workhorse of large scale simulations, trading minimal accuracy for a faster computation. We will keep improving PANNA with state of the art optimization techniques such as CUDA based featurization libraries, and support for new descriptors and improved architectures to move towards making this possibility a reality in materials modeling.



\section{Data Availability}
The data that support the findings of this study are available from the corresponding author
upon reasonable request.

\bibliography{PANNA2}
\end{document}